\documentclass[12pt]{article}
\usepackage{amssymb,amsmath,epsfig}

\begin{document}
\allowdisplaybreaks
\title{\bf Accretion Onto a Charged Higher-Dimensional Black Hole}
\author{M. Sharif \thanks{msharif.math@pu.edu.pk} and
Sehrish Iftikhar
\thanks{sehrish3iftikhar@gmail.com}~~\thanks{On leave from Department of Mathematics, Lahore College
for Women University, Lahore-54000, Pakistan.}\\
Department of Mathematics, University of the Punjab,\\
Quaid-e-Azam Campus, Lahore-54590, Pakistan.}
\date{}
\maketitle
\begin{abstract}
This paper deals with the steady-state polytropic fluid accretion
onto a higher-dimensional Reissner-Nordstr$\ddot{o}$m black hole. We
formulate the generalized mass flux conservation equation, energy
flux conservation and relativistic Bernoulli equation to discuss the
accretion process. The critical accretion is investigated by finding
critical radius, critical sound velocity and critical flow velocity.
We also explore gas compression and temperature profiles to analyze
the asymptotic behavior. It is found that the results for
Schwarzschild black hole are recovered when $q=0$ in four
dimensions. We conclude that accretion process in higher dimensions
becomes slower in the presence of charge.
\end{abstract}
\textbf{Keywords:} Accretion; Higher dimensional charged black hole.\\
\textbf{PACS:} 04.20.-q; 04.50.Gh; 04.25.dg.

\section{Introduction}

Black hole (BH) is one of the celestial objects having strong
gravitational pull that the nearby matter even light cannot escape
from its gravitational field. There are different ways to detect BH
in binary systems and center of galaxies as it cannot be observed
directly. The detection of its effect on the nearby matter is the
one way and the most promising is accretion. In astrophysics,
accretion is defined as the inward flow of matter surrounding a
compact object under the influence of gravitational field. Recent
developments in the study of quasars luminosity, relationship among
masses of massive BHs and the properties of their host galaxies
motivated the idea of accretion onto BHs \cite{1}.

Like most of the substances in the universe, all accreting matter is
in gaseous form. The problem of gas accretion by a star was first
studied by Hoyle and Lyttleton \cite{2} and later by Bondi and Hoyle
\cite{3}. The steady-state spherically symmetric accretion was
considered by Bondi \cite{4} in which star was considered at rest at
infinite gas cloud. In his classical model, he studied the flow of a
barotropic fluid in the context of Newtonian gravity. Michel
\cite{5} extended this work in the framework of GR by investigating
steady-state spherically symmetric inflow of gas onto Schwarzschild
BH. Other important studies in this context are luminosity and
frequency spectrum, the effect of magnetic field on accreting
ionized gases and accretion onto a rotating BH \cite{6}. Malec
\cite{7} investigated relativistic spherically symmetric accretion
onto a BH with and without back reaction. He found that relativistic
effects raise accretion mass in the absence of back reaction.

Shatskiy and Andreev \cite{8} studied accretion onto a non-rotating
compact object in comoving frame and explored dynamics of event
horizon formation. Jamil et al. \cite{9} analyzed the effect of
phantom like dark energy onto Reissner-Nordstr$\ddot{o}$m (RN) BH
and found that accretion is possible only through outer horizon.
Jamil and Akbar \cite{10} investigated accretion of exotic phantom
energy onto $(2+1)$-dimensional Banados-Teitelboim-Zanelli (BTZ) BH
and showed that mass accretion due to phantom energy is independent
of BH mass. Babichev et al. \cite{11} described steady-state
spherically symmetric accretion of a perfect fluid as well as scalar
field onto RN BH and found the formation of static atmosphere of
fluid around naked singularity. The same authors \cite{12} studied
accretion of spherically symmetric metric with back reaction and
showed that the metric is of Vaidya form near the horizon using
perturbation. de Freitas Pacheco \cite{13} examined relativistic as
well as non-relativistic accretion onto RN BH using two equations of
state (EOS) and found that accretion was slightly affected in the
first case while in the second case it reduced upto 60\% then the
schwarzschild BH (for extreme RN case). Sharif and Abbas \cite{14}
investigated phantom accretion onto a magnetically stringy charged
BH and found that BH does not transform into extremal BH or naked
singularity.

Recently, Park and Ricotti \cite{15} studied the increase in
luminosity and growth rates of BHs moving at super sonic speed.
Gaspari et al. \cite{16} suggested that cooling rates are tightly
linked to the BH accretion rates ($\dot{M}_{BH}\approx
\dot{M}_{cool,core}$) in the galactic core. Karkowski and Malec
\cite{17} studied steady accretion onto BH that is immersed in a
cosmological universe and found that dark energy may halt this type
of accretion. Babichev et al. \cite{18} investigated the interaction
of dark energy with Schwarzschild as well as RN BH and gave physical
reasons of decrease in mass due to accretion of phantom energy.
Ganguly et al. \cite{19} examined the process of accretion on
$4$-dimensional string cloud and found an increase in the accretion
rate with respect to string cloud parameter.

The study of gravity in a theory such as braneworld (which implies
the existence of extra dimensions) has attracted many people from
the last few decades. This theory is based on the fact that
($3+1$)-dimensional brane is embedded in a ($4+n$)-dimensional
spacetime with $n$ compact spacelike dimensions \cite{20}. It is
suggested that in braneworld theory, the effects of quantum gravity
can be observed in laboratory at TeV energies. Also, these theories
recommend that higher-dimensional BHs can be produced in large
hadron colliders and cosmic ray experiments. With the development of
higher-dimensional theories \cite{21}, it would thus be interesting
to study BHs in higher-dimensions

Tangherlini \cite{22} was the pioneer in generalization of
Schwarzschild BH in higher dimensions. Dadhich et al. \cite{23}
found the first static spherically symmetric BH solution in higher
dimensions in the context of braneworld, which has the same
structure as 4-dimensional RN BH. The physics of higher dimensional
BH is much different and richer than in 4-dimensions \cite{24}.
Accretion in higher-dimensions onto TeV-scale BHs was first studied
by Giddings and Mangano \cite{25} in Newtonian background. Sharif
and Abbas \cite{26} investigated phantom energy accretion onto a
5-dimensional charged BH and found the validity of cosmic censorship
hypothesis. John et al. \cite{27} examined steady state accretion
onto higher-dimensional Schwarzschild BH and found decrease in
accretion mass. Debnath \cite{28} studied accretion onto
higher-dimensional charged BTZ BH assuming modified Chaplygin gas as
accreting matter and found that initially BH mass increases and then
decreases to a certain finite value for phantom stage.

In this paper, we study steady-state accretion onto a
$D$-dimensional RN BH using the technique of Michel \cite{5} as well
as Shapiro and Teukolsky \cite{29}. The paper is organized as
follows: In section \textbf{2}, we study analytic relativistic
perfect fluid accretion onto RN BH in higher dimensions. Section
\textbf{3} investigates accretion on critical points. We also study
critical accretion with polytropic EOS and obtain expressions for
gas compression and temperature profile near horizon. Finally, we
summarize and discuss the results in the last section.

\section{General Formalism for Spherical Accretion}

In this section, we develop a general framework for accretion onto a
higher-dimensional spacetime and study laws of conservation of mass
and energy. The static spherically symmetric higher-dimensional RN
BH is given by \cite{30}
\begin{equation}\label{1t}
ds^{2}=-f(r)dt^{2}+f^{-1}(r)dr^{2}+r^{2}(d\Omega^{2}_{D-2}),
\end{equation}
where $D$ is the spacetime dimension and
$$d\Omega^{2}_{D-2}=d\theta^{2}_{1}+\sin^{2}\theta_{1}d\theta^{2}_{2}+
\sin^{2}\theta_{1}\sin^{2}\theta_{2}d\theta^{2}_{3}+...+
\prod^{D-3}_{\mu=1}\sin^{2}\theta_{\mu}d\theta^{2}_{D-2}$$ is the
line element on $(D-2)$-dimensional unit sphere whose volume is
\begin{equation}\nonumber
\Omega_{D-2}=\frac{2\pi^{\frac{D-1}{2}}}{\Gamma(\frac{D-1}{2})}.
\end{equation}
The lapse function in terms of mass and charge parameters $\mu$ and
$q$ is
\begin{equation}\nonumber
f(r)=1-\frac{2\mu}{r^{D-3}}+\frac{q^{2}}{r^{2(D-3)}},
\end{equation}
where $\mu=\frac{8\pi GM}{(D-2)\Omega_{(D-2)}}$ and $q=\sqrt{\frac{8
\pi G}{(D-2)(D-3)}}Q$ are the ADM mass and charge, respectively.
When $q^{2}>\mu^{2}$, this solution develops singularity at $r=0$
while for $q^{2}\leq\mu^{2}$, $f(r)$ has two real roots
\begin{equation}\nonumber r_{\pm}=(\mu\pm\mu\sqrt{1-\frac{q^{2}}
{\mu^{2}}})^{\frac{1}{(D-3)}},
\end{equation}
where $r_{+}$ is the outer horizon and $r_{-}$ is the Cauchy
horizon.

We consider steady-state inflow of gas onto central mass of BH in
radial direction. The gas is assumed to be a perfect fluid specified
by the energy-momentum tensor
\begin{equation}
T^{\sigma\upsilon}=(\rho+p)u^{\sigma}u^{\upsilon}+pg^{\sigma\upsilon}.
\end{equation}
Here $p$ and $\rho$ are the pressure and energy density of the fluid
and $u^{\sigma}=\frac{dx^{\sigma}}{ds}$ is the fluid $D$-velocity
which satisfies the normalization condition
$u^{\sigma}u_{\sigma}=-1$. We also define the baryon number flux
$J^{\sigma}=nu^{\sigma}$, where $n$ is the proper baryon number
density. The accretion process depends upon two laws of
conservation. If no particles are created or destroyed then the
particle number is conserved, i.e.,
\begin{equation}\label{1}
\nabla_{\sigma}J^{\sigma}=\nabla_{\sigma}(nu^{\sigma})=0.
\end{equation}
The law of conservation of energy-momentum tensor gives
\begin{equation}\label{2}
\nabla_{\sigma}T^{\sigma}_{\upsilon}=0.
\end{equation}
The non-zero components of D-velocity are $u^{0}=\frac{dt}{ds}$ and
$\nu(r)=u^{1}=\frac{dr}{ds}$. Using $u^{\sigma}u_{\sigma}=-1$, we
have
\begin{equation}\nonumber
u^{0}=\frac{(\nu^{2}+1-\frac{2\mu}{r^{d-3}}+\frac{q^{2}}{r^{2(d-3)}})^{\frac{1}{2}}}
{1-\frac{2\mu}{r^{D-3}}+\frac{q^{2}}{r^{2(D-3)}}}.
\end{equation}
For $D$-dimensional RN BH, Eq.(\ref{1}) takes the form
\begin{equation}\label{3}
\frac{1}{r^{D-2}}\frac{d}{dr}(r^{D-2}n\nu)=0.
\end{equation}
The null and radial components of Eq.(\ref{2}) can be written as
\begin{equation}\label{4}
\frac{1}{r^{D-2}}\frac{d}{dr}[r^{D-2}\nu(\rho+p)(\nu^{2}+1-\frac{2\mu}{r^{D-3}}+
\frac{q^{2}}{r^{2(D-3)}})^{\frac{1}{2}}]=0,
\end{equation}

\begin{equation}\label{5}
\nu\frac{d\nu}{dr}=-\left[\frac{dp}{dr}(\frac{\nu^{2}+1-\frac{2\mu}{r^{D-3}}+
\frac{q^{2}}{r^{2(D-3)}}}{\rho+p})+(D-3)(\frac{\mu}{r^{D-2}}-\frac{q^{2}}{r^{2D-5}})\right].
\end{equation}
For $q=0$ and $D=4$, the above equations reduce to the expressions
for the Schwarzschild BH \cite{5,29}.

\section{Critical Accretion}

This section is devoted to study the solutions passing through
critical points that describe the material falling into BH with
increasing velocity. The behavior of falling fluid at critical
points can express a variety of changes close to the compact
objects. The speed of sound in the medium is also very important for
a fluid as shown in the classical paper by Bondi \cite{4}. We
consider an adiabatic fluid for which there is no entropy
production, hence the law of conservation of mass energy is defined
as \cite{29}
\begin{equation}\label{5.1}
TdS=0=d(\frac{\rho}{n})+pd(\frac{1}{n}),
\end{equation}
where $S$ is the entropy per baryon and $T$ is the temperature. It
may be written as $\frac{d\rho}{dn}=\frac{\rho+p}{n}$, leading to
the adiabatic sound speed $\alpha$
\begin{equation}\label{6}
\alpha^{2}\equiv \frac{dp}{d\rho}=\frac{n}{\rho+p}\frac{dp}{d\rho}.
\end{equation}
Using this equation, the baryon and energy-momentum conservation
become
\begin{eqnarray}\label{7}
\frac{\nu'}{\nu}&+&\frac{n'}{n}+\frac{D-2}{r}=0,\\\label{8}
\nu\nu'&+&\alpha\frac{n'}{n}
(1-\frac{2\mu}{r^{D-3}}+\frac{q^{2}}{r^{2(D-3)}}+\nu^{2})+(D-3)
(\frac{\mu}{r^{D-2}}-\frac{q^{2}}{r^{2D-5}})=0,\nonumber\\
\end{eqnarray}
where prime denotes differentiation with respect to $r$. Using
Eqs.(\ref{7}) and (\ref{8}), we obtain
\begin{equation}\label{9}
\nu'=\frac{X_{1}}{X},\quad n'=\frac{X_{2}}{X},
\end{equation}
where
\begin{eqnarray}\label{64}
X_{1}&=&\frac{1}{n}\left[\frac{\alpha(D-2)}{r}(\nu^{2}+1-
\frac{2\mu}{r^{D-3}}+\frac{q^{2}}{r^{2(D-3)}})-(D-3)
(\frac{\mu}{r^{D-2}}-\frac{q^{2}}{r^{2D-5}})\right],\nonumber\\
\\\label{11}
X_{2}&=&-\frac{1}{\nu}\left[\frac{\nu^{2}(D-2)}{r}-(D-3)
(\frac{\mu}{r^{D-2}}-\frac{q^{2}}{r^{2D-5}})\right],
\\\label{12}
X&=&\frac{\nu^{2}-\alpha(\nu^{2}+1-\frac{2\mu}{r^{D-3}}+
\frac{q^{2}}{r^{2(D-3)}})}{\nu n}.
\end{eqnarray}

For large values of $r$ ($r\rightarrow\infty$), the flow satisfies
$\nu^{2}\ll1$ and is subsonic ($\nu^{2}<\alpha^{2}$) while the sound
speed must be sub-luminal ($\alpha^{2}<1$), thus Eq.(\ref{12})
implies that
\begin{equation}\label{12.1}
X\simeq\frac{\nu^{2}-\alpha^{2}}{\nu n}<0.
\end{equation}
At the event horizon, we have
\begin{equation}\label{12.2}
X=\frac{\nu^{2}(1-\alpha^{2})}{\nu n}>0,
\end{equation}
under the causality constraint $\alpha^{2}<1$. It is mentioned here
that Eq.(\ref{12.2}) is possible only for the extreme RN case, i.e.,
for $q=\mu$. From Eq.(\ref{12.1}) and (\ref{12.2}), we see that $X$
must pass through zero at $r=r_{c}$. A flow with constant energy and
entropy must be smooth at every point. Thus, if denominator vanishes
at some point, the numerator must also vanish at that point, so for
a smooth flow we must have $X_{1}=X_{2}=X=0$ at $r=r_{c}$ \cite{29}.
From Eqs.(\ref{64})-(\ref{12}), we obtain a relationship between
flow and sound velocity as
\begin{equation}\label{14}
\nu^{2}_{c}=\frac{\alpha^{2}_{c}(1-\frac{q^{2}}{r^{2(D-3)}})}
{1+\alpha^{2}_{c}(\frac{D-1}{D-3})}
=\frac{D-3}{D-2}(\frac{\mu}{r^{D-3}}-\frac{q^{2}}{r^{2(D-3)}}),
\end{equation}
where $\nu_{c}\equiv\nu(r_{c})$ and $\alpha_{c}\equiv\alpha(r_{c})$.
In the absence of charge parameter in four-dimensions, the above
relation is exactly the same as obtained in \cite{27}. To determine
the accretion rate $\dot{M}$, we integrate Eq.(\ref{3}) over a
$(D-1)$-dimensional volume and multiply by the baryon mass, $m_{b}$,
it follows that
\begin{equation}\label{15}
\dot{M}=\frac{2 \pi^{\frac{D-1}{2}}}{\Gamma(\frac{D-1}{2})}
r^{D-2}m_{b}n\nu,
\end{equation}
where $\dot{M}$ is the constant of integration (independent of $r$,
having dimension of mass per unit time), related to mass accretion
rate \cite{4}. Equation (\ref{15}) is the generalization of the
Bondi's accretion rate in higher-dimensions. For $q=0$ and $D=4$, it
reduces to the Schwarzschild case.

Following \cite{18}, Eqs.(\ref{3}) and (\ref{4}) lead to
\begin{eqnarray}\label{17}
n\nu r^{D-2}&=&A,\\\label{18}
(\frac{\rho+p}{n})^{2}(1-\frac{2\mu}{r^{D-3}}+
\frac{q^{2}}{r^{2(D-3)}})&=&(\frac{\rho_{\infty}+
p_{\infty}}{n_{\infty}})^{2},
\end{eqnarray}
where $A$ is the constant of integration. Differentiating
Eqs.(\ref{17}) and (\ref{18}) and eliminating $d\rho$, we have
\begin{equation}\label{19}
\frac{d\nu}{dr}=-\frac{\nu}{r}\frac{\left[V^{2}-\frac{\nu^{2}}
{1-\frac{2\mu}{r^{D-3}}+\frac{q^{2}}{r^{2(D-3)}}+\nu^{2}}\right]}
{\left[(D-2)V^{2}-\frac{(D-3)(\frac{\mu}{r^{D-3}}-\frac{q^{2}}{r^{2(D-3)}})}
{1-\frac{2\mu}{r^{D-3}}+\frac{q^{2}}{r^{2(D-3)}}+\nu^{2}}\right]},
\end{equation}
where $V^{2}\equiv\frac{dln(\rho+p)}{dn}-1$, which is equal to sound
velocity $\frac{dp}{d\rho}=\alpha^{2}_{c}$ . Equation (\ref{18}) is
the generalized Bernoulli equation in $D$-dimensions for a charged
BH. Equating numerator and denominator to zero, we obtain
Eq.(\ref{14}) and
\begin{equation}\label{b}
V^{2}=\alpha^{2}_{c}=\frac{(D-3)(\mu r^{D-3}-q^{2})}
{(D-2)r^{2(D-3)}-(D-1)\mu r^{D-3}+q^{2}}.
\end{equation}
Equation (\ref{b}) yields the critical radius as
\begin{equation}\label{20}
r^{D-3}_{c}=\frac{\mu
\alpha^{2}_{c}(D-1)+(D-3)}{2\alpha^{2}_{c}(D-2)}
\left[1\pm\left[1-\frac{4\alpha^{2}_{c}(D-2)
(D-3+\alpha^{2}_{c})}{\alpha^{2}_{c}(D-1)+(D-3)}
\frac{q^{2}}{\mu^{2}}\right]^{\frac{1}{2}}\right].
\end{equation}
This equation leads to two possible solutions for the critical
radius corresponding to $+$ and $-$ signs. The first indicates the
critical radius outside the event horizon which is a physically
acceptable solution. The second possibility shows the critical
radius between inner and outer horizons \cite{9}. In the present
analysis, we are interested only in the first solution. In the limit
$D=4$, our results correspond to \cite{13,18}.

\subsection{Accretion with Polytropic Equation of State}

The physical state of homogeneous substance can be described by EoS.
In order to determine an explicit value of $\dot{M}$ as well as all
the fundamental characteristics of flow, Eqs.(\ref{15}) and
(\ref{18}) must be analyzed using EoS. We consider the polytropic
EoS
\begin{equation}\label{21}
p= k n^{\omega},
\end{equation}
where $k$ is constant and $\omega$ is adiabatic
index satisfying $1<\omega<\frac{5}{3}$. Inserting Eq.(\ref{21})
into (\ref{5.1}) and integrating, we obtain
\begin{equation}\label{22}
\rho= \frac{k}{\omega-1}n^\omega+m_{b}n,
\end{equation}
where $m_{b}$ is the constant of integration obtained by comparing
with total energy density $\rho=m_{b}n+\varepsilon$, $m_{b}n$ is the
rest mass-energy density and $\varepsilon$ is the internal energy
density. From Eqs.(\ref{21}) and (\ref{22}), we have
\begin{equation}
\omega k
n^{\omega-1}=\frac{m_{b}\alpha^{2}}{1-\frac{\alpha^{2}}{\omega-1}}.
\end{equation}
When $\frac{\alpha^{2}}{\omega-1}\ll1$, we have
$n\sim\alpha^{\frac{2}{\omega-1}}$ \cite{29}, leading to
\begin{equation}
\frac{n_{c}}{n_{\infty}}\approx(\frac{\alpha_{c}}
{\alpha_{\infty}})^{\frac{2}{\omega-1}}.
\end{equation}
Using Eq.(\ref{21}) and (\ref{22}) in (\ref{18}), it follows that
\begin{equation}\label{23}
(1+\frac{\alpha^{2}}{\omega-1-\alpha^{2}})^{2}
(1-\frac{2\mu}{r^{D-3}}+\frac{q^{2}}{r^{2(D-3)}}+\nu^{2})=
(1+\frac{\alpha^{2}_{\infty}}{\omega-1-\alpha^{2}_{\infty}})^{2}.
\end{equation}
At critical radius, using (\ref{14}) and inverting Eq.(\ref{23}), we
obtain
\begin{equation}\label{24}
(1-\frac{\alpha^{2}_{c}}{\omega-1})^{2}
(\frac{(D-3)+\alpha^{2}_{c}(D-1)}{D-3})
=(1-\frac{\alpha^{2}_{\infty}}{\omega-1})^{2},
\end{equation}

For large values of $r$ ($r>r_{c}$), the baryons are expected to be
non-relativistic ($T\ll mc^{2}/\texttt{k}=10^{13}\texttt{K}$), where
$\texttt{k}$ and $\texttt{K}$ are Planck length and Kelvin,
respectively. In this system, we must have
$\alpha\ll\alpha_{c}\ll1$. Expanding Eq.(\ref{24}), we obtain a
relationship between sound velocity at critical point and the point
at infinity as
\begin{equation}\label{25}
\alpha^{2}_{c}\approx\frac{2\alpha^{2}_{\infty}(D-3)}{(3D-7)-\omega(D-1)},
\end{equation}
which corresponds to \cite{13} for $D=4$. Using Eq.(\ref{25}),
critical radius takes the form
\begin{eqnarray}\nonumber
r^{D-3}_{c}&\approx&\left[1+\left[\frac{(8\alpha^{2}_{\infty}(D-2)(D-3))
((3D-7)-\omega(D-1)+2\alpha^{2}_{\infty})}
{(2\alpha^{2}_{\infty}(D-1)-(2D-4))((2D-7)-\omega(D-1))}
\frac{q^{2}}{\mu^{2}}\right]^\frac{1}{2}\right]
\\\label{26}
&\times&\frac{\mu((3D-7)-\omega(D-1))}{4\alpha^{2}_{\infty}(D-2)(D-3)}.
\end{eqnarray}
We evaluate Bondi mass accretion rate $\dot{M}$ at the critical
point from Eq.(\ref{15}) as follows
\begin{equation}\label{28}
\dot{M}=\frac{2\pi^{\frac{D-1}{2}}}{\Gamma(\frac{D-1}{2})}
r^{D-2}_{c}m_{b}n_{c}\nu_{c}
=\frac{2\pi^{\frac{D-1}{2}}}{\Gamma(\frac{D-1}{2})}\lambda_{c}
\mu^{\frac{D-2}{D-3}}m_{b}n_{\infty}\alpha^{-(\frac{D-1}{D-3})}_{\infty}f(e),
\end{equation}
where
\begin{equation}\label{29}
\lambda_{c}=(\frac{D-3}{2})^{\frac{\omega+1}{2(\omega-1)}}
\left(\frac{(3D-7)-\omega(D-1)}{4}\right)^{\frac{7-3D+\omega(D-1)}{2(D-3)(\omega-1)}}
(D-2)^{\frac{2-D}{D-3}},
\end{equation}
is dimensionless accretion parameter and
\begin{eqnarray}\nonumber
f(e)&=&\left[1+\left[\frac{(8\alpha^{2}_{\infty}(D-2)(D-3))
((3D-7)-\omega(D-1)+2\alpha^{2}_{\infty})e^{2}}
{(2\alpha^{2}_{\infty}(D-1)-(2D-4))
((2D-7)-\omega(D-1))}\right]^{\frac{D-1}{4(D-3)}}\right]
\\\nonumber
&\times&\left[1-e^{2}\left[\frac{(3D-7)-
\omega(D-1)}{4(D-2)(D-3)\alpha^{2}_{\infty}} \left[1
\right.\right.\right.\\\label{62}&+&\left.\left.\left.
\left[\frac{(8\alpha^{2}_{\infty}(D-2)(D-3))
((3D-7)-\omega(D-1)+2\alpha^{2}_{\infty})e^{2}}
{(2\alpha^{2}_{\infty}(D-1)-(2D-4))
((2D-7)-\omega(D-1))}\right]^{\frac{1}{2}}\right]\right]^{-1}\right],\nonumber\\
\end{eqnarray}
here $e=\frac{q^{2}}{\mu^{2}}$. Re-writing Eq.(\ref{28}) in terms of
$G$ and $M$, we have
\begin{eqnarray}\nonumber
\dot{M}&=&\sqrt{\pi}[2^{\frac{4D-9}{D-3}}(D-2)^{-2(\frac{D-2}{D-3})}
(\frac{D-1}{2})^{\frac{1}{D-3}}]\\&\times&\nonumber(GM)^{\frac{D-2}{D-3}}
m_{b}n_{\infty}\alpha^{-(\frac{D-1}{D-3})}_{\infty}
(\frac{D-3}{2})^{\frac{\omega+1}{2(\omega-1)}}
\\&\times&\label{30}\left[\frac{3D-7-\omega(D-1)}{4}\right]
^{\frac{7-3D+\omega(D-1)}{2(D-3)(\omega-1)}}f(e),
\end{eqnarray}
and
\begin{eqnarray}\nonumber
f(e)&=&\left[1+\left[\frac{(8\alpha^{2}_{\infty}(D-2)(D-3))
((3D-7)-\omega(D-1)+2\alpha^{2}_{\infty})}
{(2\alpha^{2}_{\infty}(D-1)-(2D-4))
((2D-7)-\omega(D-1))}
\right.\right.\\\nonumber&\times&\left.\left.
\frac{\pi^{D-2}(D-2)Q^{2}}{2G(D-3)(\Gamma(\frac{D-1}{2})
^{2})}\right]^{\frac{D-1}{4(D-3)}}\right]
\\\nonumber
&\times&
\left[1-\frac{\pi^{D-2}(D-2)Q^{2}}{2G(D-3)(\Gamma(\frac{D-1}{2})
^{2})}\left[\frac{(3D-7)-
\omega(D-1)}{4(D-2)(D-3)\alpha^{2}_{\infty}}
\left[1\right.\right.\right.\\\nonumber&+&\left.\left.\left.\left[
\frac{(8\alpha^{2}_{\infty}(D-2)(D-3))
((3D-7)-\omega(D-1)+2\alpha^{2}_{\infty})}
{(2\alpha^{2}_{\infty}(D-1)-(2D-4)) ((2D-7)-\omega(D-1))}
\right.\right.\right.\right.\\\label{63}&\times&\left.\left.\left.\left.
\frac{\pi^{D-2}(D-2)Q^{2}}{2G(D-3)(\Gamma(\frac{D-1}{2})
^{2})}\right]^{\frac{1}{2}}\right]\right]^{-1}\right].
\end{eqnarray}
\begin{figure}\centering
\epsfig{file=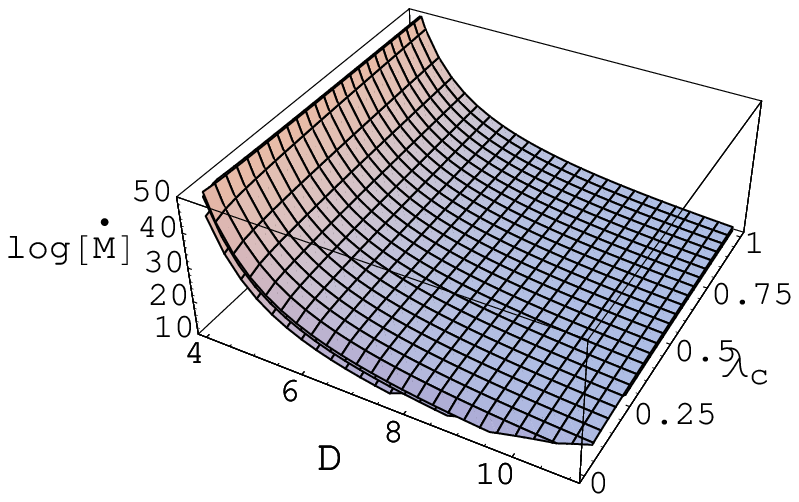,width=.44\linewidth}\epsfig{file=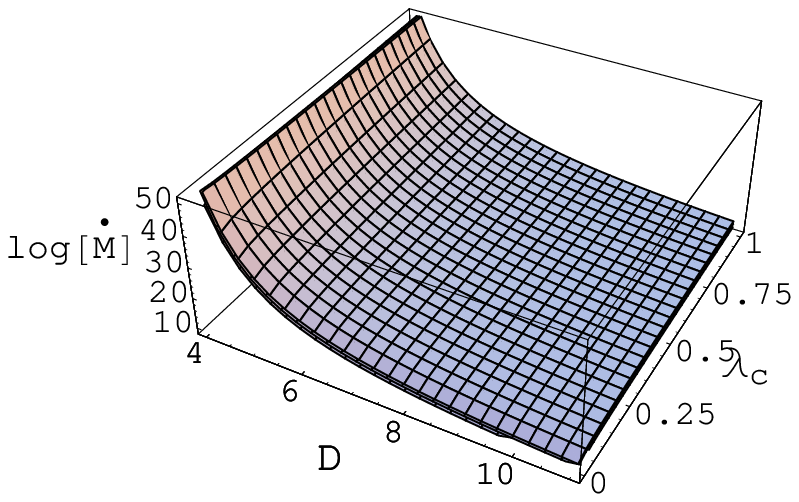,width=.44\linewidth}
\epsfig{file=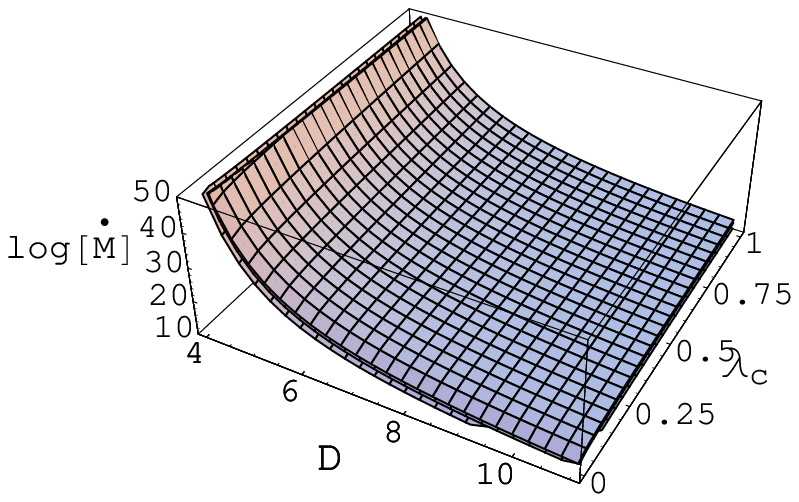,width=.44\linewidth} \caption{Plots of the
accretion rate $\dot{M}$ as a function of $D=4-11$ and $\lambda_{c}$
for $\omega$=1.1, $\frac{4}{3}$, 1.6 and $e=0.2, 0.4, 0.64$.}
\end{figure}
\begin{figure}\centering
\epsfig{file=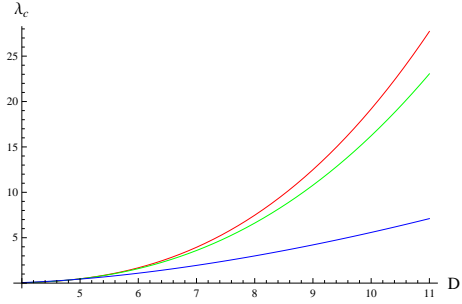,width=.44\linewidth} \caption{Behavior of
accretion parameter $\lambda_{c}$ corresponding to $D=4-11$ where
red, blue and green curves indicate $\omega$=1.1, $\frac{4}{3}$ and
1.6.}
\end{figure}

Equation (\ref{30}) shows that accretion rate in charged background
is modified by the term $f(e)$. However, the mass accretion rate
scales as $\dot{M}\sim M^{\frac{D-2}{D-3}}$ which corresponds to the
Newtonian \cite{4} as well as relativistic model \cite{29} for
$e=0,~D=4$. For the standard values of adiabatic index
($1<\omega<\frac{5}{3}$), different values of $e$ and
$G=1,~M=M_{_{\bigodot}}=1.989\times10^{33}g$, $m_{b}=
1.67\times10^{-24}g$, $\alpha_{\infty}=10^{6}cms^{-1}$,
$n_{\infty}=1 cm^{-3}$, the behavior of $r_{c},~\lambda_{c}$ and
$\dot{M}$ is given in Tables \textbf{1}-\textbf{7}. The graphical
representation of $\dot{M}$ and $\lambda_{c}$ is shown in Figures
\textbf{1} and \textbf{2}. It is seen that dimensions as well as
charge parameter affect the rate of accretion. The accretion rate
becomes slower as the dimension increases. The rate of accretion for
small values of charge is higher as compared to large values. Thus
$\dot{M}$ shows decreasing behavior for increasing dimensions as
well as charge.\\\\
\textbf{Table 1:} Accretion parameter $\lambda_{c}$ for
$\omega=1.2,~\frac{4}{3},~1.6$.
\begin{table}[bht]
\centering
\begin{small}
\begin{tabular}{|c|c|c|c|c|}
\hline\textbf{$D$} &\textbf{$\lambda_{c1}$} &\textbf{$\lambda_{c2}$}
&\textbf{$\lambda_{c3}$}\\
\hline4&0.2488&0.1768&0.0917\\
\hline5&0.4967&0.4330&0.3545\\
\hline6&0.6004&0.5473&0.4818\\
\hline7&0.6543&0.6077&0.5503\\
\hline8&0.6868&0.6444&0.5922\\
\hline9&0.7085&0.6689&0.6202\\
\hline10&0.7239&0.6864&0.6403\\
\hline11&0.7353&0.6994&0.6553\\
\hline
\end{tabular}
\end{small}
\end{table}\\

\subsection{Asymptotic Analysis}

Here we estimate the flow parameters for $r_{H}<r\ll r_{c}$ as well
as $r=r_{c}$. The gas passes through supersonic flow at distance
below Bondi radius, i.e., $\nu>\alpha$ when $r_{H}<r\ll r_{c}$. We
find an upper bound of the radial dependence of gas velocity
\cite{27,29}
\begin{equation}\label{31}
\nu^{2}\approx\frac{2\mu}{r^{D-3}}-\frac{q^{2}}{r^{2(D-3)}}.
\end{equation}
The gas compression rate from Eqs.(\ref{15}), (\ref{28}) and
(\ref{31}) becomes
\begin{equation}\label{32}
\frac{n(r)}{n_{\infty}}\approx\frac{\lambda_{c}}{\sqrt{2}}f(e)
(\frac{\mu}{r^{D-3}\alpha^{2}_{\infty}})^{\frac{D-1}{2(D-3)}}
(1-\frac{q^{2}}{4\mu r^{D-3}}).
\end{equation}\\\\\\
\textbf{Table 2:} Critical radius $r_{c}$ for $\omega=1.1$ and
$e=0.2,~0.4,~0.67$.
\begin{table}[bht]
\centering
\begin{small}
\begin{tabular}{|c|c|c|c|c|}
\hline\textbf{$D$} &\textbf{$r_{c1}$} &\textbf{$r_{c2}$}
&\textbf{$r_{c3}$}\\
\hline4&$9.8156\times10^{26}$&$1.9631\times10^{26}$&$3.2882\times10^{26}$\\
\hline5&$3.4519\times10^{26}$&$6.9118\times10^{26}$&$1.1577\times10^{26}$\\
\hline6&$1.7840\times10^{25}$&$3.5681\times10^{25}$&$5.9765\times10^{25}$\\
\hline7&$1.0928\times10^{25}$&$2.1855\times10^{24}$&$3.6607\times10^{25}$\\
\hline8&$7.3908\times10^{25}$&$1.4782\times10^{24}$&$2.4759\times10^{24}$\\
\hline9&$5.9955\times10^{25}$&$1.0671\times10^{24}$&$1.7874\times10^{24}$\\
\hline10&$3.4519\times10^{24}$&$8.0682\times10^{23}$&$1.3515\times10^{23}$\\
\hline11&$3.4519\times10^{24}$&$6.3162\times10^{23}$&$1.0580\times10^{23}$\\
\hline
\end{tabular}
\end{small}
\end{table}
\\\\\\
\textbf{Table 3:} Critical radius $r_{c}$ for $\omega=\frac{4}{3}$
and $e=0.2,~0.4,~0.67$.
\\\\
\begin{table}[bht]
\centering
\begin{small}
\begin{tabular}{|c|c|c|c|c|}
\hline\textbf{$D$} &\textbf{$r_{c1}$} &\textbf{$r_{c2}$}
&\textbf{$r_{c3}$}\\
\hline4&$9.1026\times10^{26}$&$1.8205\times10^{26}$&$3.0494\times10^{26}$\\
\hline5&$3.2566\times10^{26}$&$6.5133\times10^{26}$&$1.0910\times10^{26}$\\
\hline6&$1.6925\times10^{26}$&$3.3850\times10^{25}$&$5.6698\times10^{25}$\\
\hline7&$1.0405\times10^{25}$&$2.0810\times10^{25}$&$3.4857\times10^{25}$\\
\hline8&$7.0539\times10^{25}$&$1.4108\times10^{25}$&$2.3630\times10^{24}$\\
\hline9&$3.8662\times10^{25}$&$1.0201\times10^{24}$&$1.7087\times10^{24}$\\
\hline10&$3.0523\times10^{24}$&$7.1221\times10^{24}$&$1.2935\times10^{23}$\\
\hline11&$3.0351\times10^{24}$&$6.60505\times10^{23}$&$1.0135\times10^{23}$\\
\hline
\end{tabular}
\end{small}
\end{table}
\\\\\\\\\\\\\\\\
\textbf{Table 4:} Critical radius $r_{c}$ for $\omega=1.6$ and
$e=0.2,0.4,0.67$.
\begin{table}[bht]
\centering
\begin{small}
\begin{tabular}{|c|c|c|c|c|}
\hline\textbf{$D$} &\textbf{$r_{c1}$} &\textbf{$r_{c2}$}
&\textbf{$r_{c3}$}\\
\hline4&$8.4515\times10^{26}$&$1.6903\times10^{26}$&$2.8313\times10^{26}$\\
\hline5&$3.0663\times10^{26}$&$6.1326\times10^{26}$&$1.0272\times10^{26}$\\
\hline6&$1.6033\times10^{26}$&$3.2067\times10^{25}$&$5.6117\times10^{25}$\\
\hline7&$1.8910\times10^{26}$&$1.9782\times10^{25}$&$3.4618\times10^{25}$\\
\hline8&$6.7201\times10^{25}$&$1.3440\times10^{24}$&$2.3520\times10^{24}$\\
\hline9&$4.8665\times10^{25}$&$9.7330\times10^{24}$&$2.6183\times10^{24}$\\
\hline10&$3.6882\times10^{24}$&$7.3764\times10^{23}$&$1.8443\times10^{23}$\\
\hline11&$2.8922\times10^{24}$&$5.7844\times10^{23}$&$1.4461\times10^{23}$\\
\hline
\end{tabular}
\end{small}
\end{table}\\\\\\
\textbf{Table 5:} Accretion rate $\dot{M}$ for $\omega=1.1$ and
$e=0.2,~0.4,~0.67$.
\begin{table}[bht]
\centering
\begin{small}
\begin{tabular}{|c|c|c|c|c|}
\hline\textbf{$D$} &\textbf{$\dot{M}_{1}$} &\textbf{$\dot{M}_{2}$}
&\textbf{$\dot{M}_{3}$}\\
\hline4&$3.5568\times10^{40}$&$6.2876\times10^{39}$&$1.9883\times10^{37}$\\
\hline5&$2.0451\times10^{27}$&$5.1129\times10^{26}$&$5.1129\times10^{24}$\\
\hline6&$9.7005\times10^{22}$&$2.7221\times10^{22}$&$3.9958\times10^{20}$\\
\hline7&$6.9321\times10^{22}$&$2.0611\times10^{20}$&$3.6657\times10^{18}$\\
\hline8&$3.4150\times10^{19}$&$3.5568\times10^{19}$&$2.0977\times10^{17}$\\
\hline9&$4.1980\times10^{18}$&$1.3223\times10^{18}$&$2.8499\times10^{16}$\\
\hline10&$8.3930\times10^{17}$&$2.6878\times10^{17}$&$6.1199\times10^{15}$\\
\hline11&$2.2132\times10^{16}$&$7.1759\times10^{16}$&$1.7028\times10^{15}$\\
\hline
\end{tabular}
\end{small}
\end{table}\\\\\\\\\\\\\\\\\\\\\\\\
\textbf{Table 6:} Accretion rate $\dot{M}$ for $\omega=\frac{4}{3}$
and $e=0.2,~0.4,~0.67$.
\begin{table}[bht]
\centering
\begin{small}
\begin{tabular}{|c|c|c|c|c|}
\hline\textbf{$D$} &\textbf{$\dot{M}_{1}$}&\textbf{$\dot{M}_{2}$}
&\textbf{$\dot{M}_{3}$}\\
\hline4&$7.0978\times10^{37}$&$1.2547\times10^{37}$&$1.9559\times10^{38}$\\
\hline5&$1.6580\times10^{25}$&$4.1449\times10^{24}$&$3.7304\times10^{25}$\\
\hline6&$1.2230\times10^{21}$&$3.4321\times10^{20}$&$2.5720\times10^{21}$\\
\hline7&$1.0816\times10^{19}$&$3.2273\times10^{18}$&$2.0681\times10^{19}$\\
\hline8&$6.0816\times10^{18}$&$1.8721\times10^{17}$&$1.2116\times10^{18}$\\
\hline9&$1.7298\times10^{17}$&$2.5651\times10^{16}$&$1.6003\times10^{17}$\\
\hline10&$3.5541\times10^{16}$&$5.5404\times10^{15}$&$3.3367\times10^{16}$\\
&$4.7735\times10^{16}$&$1.5481\times10^{15}$&$9.2243\times10^{15}$\\
\hline
\end{tabular}
\end{small}
\end{table}
\\\\\\
\textbf{Table 7:} Accretion rate $\dot{M}$ for $\omega=1.6$ and
$e=0.2,~0.4,~0.67$.
\begin{table}[bht]
\centering
\begin{small}
\begin{tabular}{|c|c|c|c|c|}
\hline\textbf{$D$} &\textbf{$\dot{M}_{1}$} &\textbf{$\dot{M}_{2}$}
&\textbf{$\dot{M}_{3}$}\\
\hline4&$1.8221\times10^{39}$&$1.8438\times10^{38}$&$8.9820\times10^{37}$\\
\hline5&$3.1419\times10^{26}$&$5.0271\times10^{25}$&$2.8278\times10^{25}$\\
\hline6&$1.2230\times10^{21}$&$3.4321\times10^{20}$&$2.5720\times10^{21}$\\
\hline7&$1.5512\times10^{20}$&$3.1211\times10^{19}$&$1.8865\times10^{19}$\\
\hline8&$8.1671\times10^{18}$&$1.7203\times10^{18}$&$1.0549\times10^{18}$\\
\hline9&$1.0480\times10^{18}$&$2.2760\times10^{17}$&$1.4092\times10^{17}$\\
\hline10&$2.1592\times10^{17}$&$4.7927\times10^{16}$&$2.9878\times10^{16}$\\
\hline11&$5.8212\times10^{16}$&$1.3135\times10^{16}$&$8.2307\times10^{15}$\\
\hline
\end{tabular}
\end{small}
\end{table}\\
We consider a Maxwell-Boltzmann gas, $P=nk_{\texttt{B}}T$. From
Eqs.(\ref{21}) and (\ref{32}) we calculate the adiabatic temperature
profile as
\begin{equation}\label{33.7}
\frac{T(r)}{T_{\infty}}=\frac{n(r)}{n_{\infty}}\approx
\left[\frac{\lambda_{c}}{\sqrt{2}}f(e)
(\frac{\mu}{r^{D-3}\alpha^{2}_{\infty}})^{\frac{D-1}{2(D-3)}}
(1-\frac{q^{2}}{4\mu r^{D-3}})\right]^{\omega-1},
\end{equation}
at the event horizon $r=r_{H}=[\mu(1+(1-e^{2})^{\frac{1}{2}})]
^{\frac{1}{D-3}}$. Since the flow is supersonic below the Bondi
radius, so the flow velocity is still approximated by Eq.(\ref{31}).
At the event horizon, we have
$\nu_{H}=\nu^{2}(r_{H})\approx(1-\frac{e^{2}}{1+(1-e^{2})^{\frac{1}{2}}})
\frac{1}{1+(1-e^{2})^{\frac{1}{2}}}$. Using Eqs.(\ref{32}) and
(\ref{33.7}), the gas compression rate and the adiabatic temperature
profile at the event horizon take the following form
\begin{eqnarray}\label{34}
\frac{n_{H}}{n_{\infty}}&\approx&\frac{\lambda_{c}f(e)}
{\sqrt{2}(1+(1-e^{2})^{\frac{D-1}{2(D-3)}})}
(\frac{4(1+(1-e^{2})^{\frac{1}{2}})-e^{2}}
{4(1+(1-e^{2})^{\frac{1}{2}})})
(\frac{c}{\alpha_{\infty}})^{\frac{D-1}{D-3}},
\\\label{35}
\frac{T_{H}}{T_{\infty}}&\approx&\left[\frac{\lambda_{c}f(e)}
{\sqrt{2}(1+(1-e^{2})^{\frac{D-1}{2(D-3)}})}
(\frac{4(1+(1-e^{2})^{\frac{1}{2}})-e^{2}}
{4(1+(1-e^{2})^{\frac{1}{2}})})
(\frac{c}{\alpha_{\infty}})^{\frac{D-1}{D-3}}\right]^{\omega-1},
\end{eqnarray}
where $c$ is the speed of light. In four-dimensional case, when
$e=0$ the above expressions correspond to the spherical accretion
onto Schwarzschild BH \cite{29}.

\section{Concluding Remarks}

It is believed that matter accreting onto a gravitating body is the
source of power supply in closed binary systems, galactic nuclei and
quasars \cite{31}. There has been a growing interest to study
theories which predict gravity in extra dimensions such as string
theories and braneworld cosmology. This paper provides the effect of
steady-state spherically symmetric adiabatic accretion onto a
charged $D$-dimensional BH and explores critical accretion following
Michel \cite{5} as well as Shapiro and Teukolsky \cite{29}. The
critical radius and mass accretion rate as well as the gas
compression and temperature profile (below the critical radius and
at the event horizon) are found. It turns out that mass accretion
rate depends upon BH mass and dimensions. Also, $\dot{M}$ is
modified by the term $f(e)$ which continuously decreases as the
dimension increases and the accretion rate for large values of
charge is less than that of small values. We observe that accretion
rate decreases gradually but the process is slower than the
higher-dimensional Schwarzschild BH \cite{27}. We conclude that the
accretion rate of charged BH slows down in higher dimensions. It is
interesting to mention here that all our results for $q=0$ and $D=4$
correspond to accretion rate of Schwarzschild BH. This leads to the
generalization of the results presented in \cite{5,29} in terms of
accretion onto a charged BH in higher-dimensions.


\begin{thebibliography}{43}


\bibitem{1} Ho, L.C.: \textit{Coevolution of Black Holes and Galaxies} (Cambridge, 2004).

\bibitem{2} Hoyle, F. and Lyttleton, R.A.: Proc. Cambridge Philos. Soc. \textbf{35}(1939)405.

\bibitem{3} Bondi, H. and Hoyle, F.: Mon. Not. R. Astron. Soc. \textbf{104}(1944)273.

\bibitem{4} Bondi, H.: Mon. Not. R. Astron. Soc. \textbf{112}(1952)195.

\bibitem{5}  Michel, F.C.: Astrophys. Space Sci. \textbf{15}(1972)153.

\bibitem{6} Shapiro, S.L.: Astrophys. J. \textbf{ 180}(1973)531; ibid. \textbf{185}(1973)69; ibid. \textbf{189}(1974)343.

\bibitem{7} Malec, E.:  Phys. Rev. D \textbf{60}(1999)104043.

\bibitem{8} Shatskiy, A.A. and  Andreev, A.Y. : Zh. Eksp. Teor. Fiz. \textbf{116}(1999)353.

\bibitem{9} Jamil, M., Qadir, A. and Rashid, M.A.: Eur. Phys. J. C \textbf{58}(2008)325.

\bibitem{10} Jamil, M. and Akbar, M.: Gen. Relativ. Gravit. \textbf{43}(2011)1061.

\bibitem{11} Babichev, E., Dokuchaev, V. and Eroshenko, Y.: J. Exp. Theor. Phys. \textbf{112}(2011)784.

\bibitem{12} Babichev, E., Dokuchaev, V. and Eroshenko, Y.: Class. Quantum Gravt. \textbf{29}(2012)115002.

\bibitem{13} de Freitas Pacheco, J.A.: J. Thermodyn. \textbf{2012}(2012)791870.

\bibitem{14} Sharif, M. and Abbas, G.: Chin. Phys. Lett. \textbf{29}(2012)010401.

\bibitem{15} Park, K. and Ricotti, M.: Astrophys. J. \textbf{767}(2013)163.

\bibitem{16} Gaspari, M., Ruszkowski, M. and Oh, S.P. : Mon. Not. R. Astron. Soc. \textbf{432}(2013)3401.

\bibitem{17} Karkowski, J. and Malec, E.: Phys. Rev. D \textbf{87}(2013)044007.

\bibitem{18} Babichev, E., Dokuchaev, V. and Eroshenko, Y.: Phys. Usp. \textbf{56}(2013)1155.

\bibitem{19}  Ganguly, A., Ghosh, S.G. and Maharaj, S.D: Phys. Rev. D \textbf{90}(2014)064037.

\bibitem{20}  Randall, L., Sundrum, R.: Phys. Rev. Lett. \textbf{83}(1999)3370.

\bibitem{21}  Emparan, R., Horowitz, G.T. and Myers, R.C.: Phys. Rev. Lett. \textbf{85}(2000)499.

\bibitem{22}  Tangherlini, F.R.: Nuovo Cimento \textbf{27}(1963)636 .

\bibitem{23}  Dadhich, N., Maartens, R., Papadopoulos, P. and Rezania, V.: Phys. Lett. B \textbf{487}(2000)1.

%\bibitem{24}  Kanti, P: Int. J. Mod. Phys. A \textbf{19}(2004)4899.

\bibitem{24}  Emparan, R. and Reall, H.S.: Living Rev. Rel. \textbf{11}(2008)6.

\bibitem{25}  Giddings, S.B. and Mangano, M.L.: Phys. Rev. D \textbf{78}(2008)035009.

\bibitem{26} Sharif, M. and Abbas, G.: Mod. Phys. Lett. A \textbf{26}(2011)1731.

\bibitem{27} John, A.J., Ghosh, S.G. and Maharaj, S.D.: Phys. Rev. D
\textbf{88}(2013)104005.

\bibitem{28} Debnath, U.: Eur. Phys. J. C \textbf{75}(2015)449.

\bibitem{29} Shapiro, S.L. and Teukolsky, S.A.: \textit{Black Holes, White Dwarfs
and Neutron Stars} (Wiley, 1983).

\bibitem{30} Aman, J.E. and Pidokrajt, N.: Phys. Rev. D \textbf{73}(2006)024017.

\bibitem{31} Frank, J., King, A. and Raine, D.: \textit{Accretion Power in
Astrophysics} (Cambridge University Press, 2002).



\end{thebibliography}
\end{document}